# Eclipses in Australian Aboriginal Astronomy

Duane W. Hamacher[1] and Ray P. Norris[1]

[1]Department of Indigenous Studies, Macquarie University, NSW, 2109, Australia

Corresponding Author e-mail: duane.hamacher@mq.edu.au

### Abstract

We explore 50 Australian Aboriginal accounts of lunar and solar eclipses to determine how Aboriginal groups understood this phenomenon. We summarise the literature on Aboriginal references to eclipses, showing that many Aboriginal groups viewed eclipses negatively, frequently associating them with bad omens, evil magic, disease, blood and death. In many communities, Elders or medicine men were believed to have the ability to control or avert eclipses by magical means, solidifying their role as provider and protector within the community. We also show that many Aboriginal groups understood the motions of the sun-earth-moon system, the connection between the lunar phases and tides, and acknowledged that solar eclipses were caused by the moon blocking the sun.

## 1.0    Introduction

Aboriginal Australians were careful observers of the night sky and possessed a complex understanding of the motions of celestial bodies and their correlation with terrestrial events, such as the passage of time, the changing of seasons, and the emergence of particular food sources (e.g. Haynes, 1992a, 1992b; Johnson, 1998; Fredrick, 2008; Norris & Norris, 2009). Aboriginal people used the sky for navigation, marriage and totem classes, and cultural mnemonics (Johnson, 1998). The celestial world was an important and integral aspect of the landscape, which was inseparable from the terrestrial world.  Aboriginal knowledge was passed through successive generations through oral tradition, dance, ceremony, and various artistic forms, including as paintings, drawings, and petroglyphs.  Much of this knowledge was restricted to particular genders or totems, or was dependant on the initiation that individual into the higher ranks of the community.

As part of our continuing research into Aboriginal Astronomy (Norris & Hamacher, 2009, Norris & Norris, 2009), specifically regarding transient celestial phenomena (e.g. Hamacher & Frew, 2010; Hamacher & Norris, 2009, 2010, 2011), this paper explores Aboriginal knowledge of solar and lunar eclipses.  We do this to gain a better understanding of Aboriginal sky knowledge and to determine the methods of scientific deduction from an Indigenous perspective.

Many Aboriginal cultures were heavily damaged by colonisation, and a significant amount of traditional (i.e. pre-colonisation) knowledge about celestial phenomena has been lost.  Most of the records available in the literature are colonist accounts – few of which come from professional ethnographers.  Given that Aboriginal societies are extremely complex and exist in a framework that is foreign to most Westerners, we acknowledge our limitations in interpreting the available information, which is strongly influenced on the biases, interpretations, and legitimacy of the sources.  The sources from which we draw information include traditional Aboriginal custodians and elders, Western professional researchers, and amateurs with little or no training in the recording or interpretation of Indigenous knowledge.

In this paper, we examine five aspects of traditional Aboriginal knowledge regarding eclipses: 1) Aboriginal perceptions and reactions to eclipses, 2) Aboriginal explanations regarding the causes of eclipses, 3) dating oral traditions using historic eclipses, 4) predicting eclipses, and 5) representations of eclipses in Aboriginal rock art.  We begin by discussing the science of lunar phases, tides, and eclipses.  If the account describes or is attributed to a known historic eclipse, it is given an "Event #", with the details of each event listed in Table 1 (solar and lunar eclipse data calculated using Espenak & O'Byrne, 2007a and 2007b, respectively).  We also include, in Table 2, the nomenclature origins of Australian placenames that include the word "eclipse", even if they have no direct link to Aboriginal culture.

Table 1: *Eclipses discussed in this paper are given in this table, which includes the event number (#), the date of the eclipse (DD/MM/YEAR), coordinates of the location where it was seen, the eclipse type (T: S = solar, L = lunar) and subtype (ST: P = partial, T = total, A = annular), the percentage of the sun's area eclipsed (Obs, only for solar eclipses), and the time of maximum eclipse (t, in local time). Events 2 and 8 are of the same eclipse seen from two different locations. Data are calculated using Espenak and O'Byrne (2007a, 2007b) with the following time zone conversions: WA = UTC +8:00; NT/SA = UTC +9:00; QLD/NSW/ VIC/TAS = UTC+10:00 (Eucla, WA = UTC +8:45).*

| # | Date | Location (Lat, Lon) | T | ST | Obs | t |
|---|------|---------------------|---|----|----|---|
| 1 | 30/07/1916 | 27° 20′ S, 126° 10′ E | S | A | 94.3 | 09:46:19 |
| 2 | 21/09/1922 | 25° 11′ S, 133° 11′ E | S | T | 100 | 15:55:00 |
| 3 | 13/08/1859 | 34° 55′ S, 138° 35′ S | L | T | — | 02:04 |
| 4 | 23/06/I899 | 16° 58′ S, 122° 39′ E | L | T | — | 22:18 |
| 5 | 28/12/1917 | 30° 27′ S, 131° 50′ E | L | T | — | 18:40 |
| 6 | 21/09/1922 | 32° 07′ S, 133° 40′ E | S | P | 75.7 | 14:56:37 |
| 7 | 05/04/1856 | 19° 15′ S, 146° 49′ E | S | P | 92.7 | 17:05:31 |
| 8 | 21/09/1922 | 28° 33′ S, 150° 19′ E | S | T | 100 | 16:13:26 |
| 9 a | 12/03/1793 | 31° 07′ S, 138° 23′ E | S | P | 92.9 | 15:58:26 |
| 9 b | 07/10/1782 | 31° 07′ S, 138° 23′ E | S | T | 100 | 09:21:58 |
| 10 | 22/11/1900 | 16° 58′ S, 122° 39′ E | S | P | 73.1 | 14:14:42 |
| 11 | 08/08/1831 | 33° 52′ S, 151° 13′ E | S | P | 87.6 | 07:03:55 |
| 12 | 03/10/1819 | 13° 54′ S, 126° 18′ E | L | T | — | 23:13 |
| 13 | 12/05/1873 | 22° 20′ S, 131° 38′ E | L | T | — | 20:20 |
| 14 | 12/12/1871 | 18° 46′ S, 146° 33′ E | S | P | 18.5 | 14:15:47 |
| 15 | 28/09/1791 | 35° 10′ S, 117° 53′ E | S | P | 92.2 | 06:38:46 |

Table 2: *The nomenclature behind placenames in Australia that include the word "eclipse". We were unable to locate any references that explain the nomenclature behind two locations in Western Australia with the name "Eclipse Hill" – one near Buraminya (~750 km east of Perth), and the other near Lennard Brook (~70 km north of Perth). The local Aboriginal name of Eclipse Island, Queensland is "Garoogubbee" (Bindloss 2002:330).*

| Name | State | Coordinates | Event # | Reference |
|------|-------|-------------|---------|-----------|
| Eclipse Hill | WA | 13° 54′S, 126° 18′E | 12 | Feekan et al (1970:230) |
| Eclipse Islands | WA | 13° 54′S, 126° 18′E | 12 | Feekan et al (1970:230) |
| Mount Eclipse | NT | 22° 20′S, 131° 38′E | 13 | Feekan et al (1970:164) |
| Eclipse Island | QLD | 18° 46′S, 146° 33′E | 14 | Reed (1973:87) |
| Eclipse Island | WA | 35° 10′S, 117° 53′E | 15 | Martin (1943); Reed (1973:87) |

## 2.0     The Science of the Earth-Moon-Sun System

### 2.1     Lunar Phases

As the moon orbits the earth, an earth-bound observer will see a different percentage of the moon illuminated by the sun throughout a lunar month, which are referred to as lunar phases.  These phases are divided into new, crescent, first quarter, waxing gibbous, full, waning gibbous, last quarter, crescent, back to new moon (see Figure 1).  When the moon is between the earth and sun, appearing near the sun in the sky from an earthbound perspective, it is essentially invisible to us for about three days, which we call the *new moon*.  As the moon moves towards solar opposition, more of the surface is illuminated by the sun.  When less than half of the moon is illuminated, it is called *crescent*, while more than half illuminated is called *gibbous*.  As more of the moon's surface becomes progressively illuminated, we deem it *waxing*.  When the moon is at solar opposition, the entire hemisphere of the moon facing the earth is illuminated, revealing a *full moon*.  As the moon fades, it is deemed *waning*.  The moon rises at dawn during new moon and dusk during full moon, with the first quarter moon rising at midday and the last quarter moon rising at midnight.

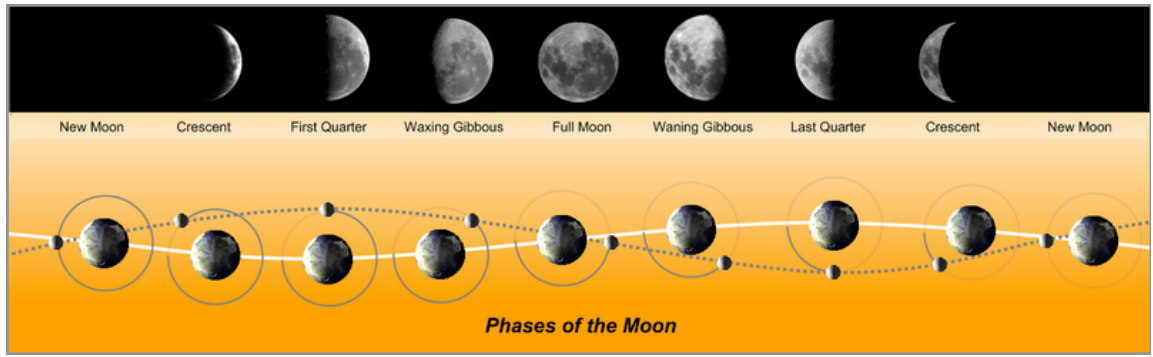

*Figure 1: Lunar phases as seen from the earth (top) and from above earth (bottom), with the sun to the left. This image corrected from the original for observers in the Southern Hemisphere. Image reproduced under Wikimedia commons licence agreement.*

To understand the causes of eclipses, it is essential to understand the relative motions of the sun and moon, which cause lunar phases. By examining Aboriginal oral traditions, we can determine if Aboriginal people in traditional times understood the relative motions of the moon-sun system and their correlation to events on the earth, such as tides.

### 2.2 Eclipses

In the earth-moon-sun system, there are two general types of eclipses: solar and lunar. When the moon passes between the earth and sun, an observer in the area on the earth that falls into the moon's shadow sees a solar eclipse. During a total solar eclipse, the sun is completely blocked and day turns completely into night (called totality). During totality, the sun's faint corona, as well as solar flares and prominences, may be observed. The presence of the corona and its shape and intensity depend on the presence of sunspots, which relate to the 11-year solar cycle (c.f. Aschwanden, 2004). Total solar eclipses are rare, and can be seen from a given point on the Earth's surface only once every 410 years, while total solar eclipses in the Southern Hemisphere are even more rare, occurring only once every 540 years (Steel, 1999:351). If only part of the sun is covered, we see a partial solar eclipse. While total eclipses are quite rare, partial eclipses are far more frequent, with > 30 such events occurring every century. If the moon eclipses the sun during apogee (due to its eccentric orbit), the moon will completely fit within the disc of the sun, leaving a ring of the solar disc visible, called the annulus. Thus, this is referred to as an annular eclipse.

There has been some debate regarding the visibility of partial eclipses. Even when 99% of the sun is eclipsed, the remaining 1% is bright enough to cause damage to the eye (Chou, 1981; March, 1982). There have been no studies that suggest what magnitude would be required for people to notice a partial eclipse, but Stephenson (1978:39) claims that partial eclipses that cover 98% of the sun's surface could go by unnoticed, unless they were known in advance, used an observing aide, or were low on the horizon and/or the light intensity was reduced by the presence of clouds (e.g. Newton, 1979: 101). Mostert (1988) claims that no unambiguous hard evidence exists that a partial eclipse has been observed with the naked eye. If this were true, would we expect accounts of solar eclipses in Aboriginal oral traditions? We determine the frequency of total solar eclipses over a 1000-year period from 900-1900 CE for 11 locations across Australia (see Table 3). An average rate of 2.36 observed total eclipses over Australia over 1000 years is roughly consistent with the estimate of Steel (1999), or approximately one every 400-500 years. Assuming a lifespan of 50 years, a total solar eclipse would be seen once every 10 generations. Given this, we would expect to find very few accounts of solar eclipses, either partial or solar.

When the moon passes through the shadow of the earth, we witness a lunar eclipse. Because the earth is so much larger than the moon, total lunar eclipses are visible from a much wider area of the earth and are far more frequent than solar eclipses, often occurring more than once per year. During a total lunar eclipse, longer wavelengths of light from the sun are refracted through the earth's atmosphere, causing the moon to take on a ruddy appearance, although the colour varies from red to orange, pink or copper. This phenomenon was noted by some Aboriginal groups.

Table 3: The frequency of total solar eclipses as seen from 11 different locations across Australia between 900-1900 CE. Information includes the name of the observation location, the number of total eclipse events ($N_e$) and the years those eclipses were observed. Data are taken from Espenak & O'Byrne (2007a).

| City | State | $N_e$ | Years of Total Eclipses |
|------|-------|-------|-------------------------|
| Alice Springs | Northern Territory | 0 | |
| Adelaide | South Australia | 5 | 1033, 1339, 1517, 1728, 1802 |
| Brisbane | Queensland | 4 | 1134, 1308, 1554, 1831 |
| Canberra | Australian Capital Territory | 1 | 1247 |
| Darwin | Northern Territory | 4 | 1191, 1242, 1256 |
| Hobart | Tasmania | 3 | 909, 1064, 1728 |
| Melbourne | Victoria | 2 | 1008, 1782 |
| Perth | Western Australia | 1 | 1310 |
| Cairns | Queensland | 0 | |
| Broome | Western Australia | 2 | 1712, 1737 |
| Cobar | New South Wales | 4 | 1308, 1336, 1547, 1608 |

## 3.0    Aboriginal Oral Traditions of the Sun and Moon

In most Aboriginal cultures, the sun is female and the moon is male (Haynes, 1992:130; Johnson, 1998), although this is not universal (e.g. Meyer, 1846:11–12). While the specific details vary between groups, many Aboriginal communities describe a dynamic between the sun and moon, typically involving one pursuing the other across the sky from day to day, occasionally meeting during an eclipse (Parker, 1905:139-140; Johnson, 1998:129; see next section). Many stories explain why the moon gets progressively "fatter" as it waxes from new moon to full moon, then fades away to nothing as it wanes back to new moon. For example, the full moon is a fat, lazy man called *Ngalindi* to the Yolngu of Arnhem Land. His wives punish his laziness (or, in some versions, his breaking of taboos) by chopping off bits of him with their axes, causing the waning moon. He manages to escape by climbing a tall tree to follow the Sun, but is mortally wounded, and dies (new moon). After remaining dead for three days, he rises again, growing fat and round (waxing moon), until his wives attack him again in a cycle that repeats to this day (Wells, 1964; Hulley, 1996).

Because the lunar month is roughly the same length as the menstrual cycle, the moon is sometimes associated with fertility, sexual intercourse, and childbearing. In some communities, young women were warned about gazing at the moon for fear of becoming pregnant (Haynes, 1997:107). The Ngarrindjeri of Encounter Bay, South Australia saw the moon as a promiscuous woman (Meyer, 1846:11-12) who became thin and wasted away (waning moon) as a result of her numerous sexual encounters. When she became very thin (crescent moon), the creator being Nurrunderi ordered her to be driven away. She was gone for a short while (new moon), but began to eat nourishing roots, causing her to fatten again (waxing moon). A similar account is given by the nearby Jaralde people, except the waxing moon represents the moon-woman coming to term in pregnancy (Berndt et al, 1993:232-233). Several other Aboriginal groups associate the moon with love, fertility and intercourse, including the Koko-Yalanyu of the Bloomfield River, Queensland (McConnell, 1931) and the Lardil people of Mornington Island in the Gulf of Carpentaria (Isaacs, 1980:163-166; Roughsey, 1971:82-84; see Johnson, 1998 and Fredrick, 2008:102-104 for more examples).

The moon and the sun have a gravitational influence on the ocean, causing tides. Higher tides than normal (spring tides) occur when the sun and moon are aligned or opposed while lower tides than normal (neap tides) occur when the sun and moon are at 90° to the earth, damping each other's gravitational influence. Many coastal groups understand the relationship between lunar phases and the ocean tides, including the correlation between the spring tide and full moon. According to the Yolngu of Arnhem Land and the Anindilyakwa of Groote Eylandt (Hulley, 1996), when the tides are high, the water fills the moon as it rises at dawn and dusk (full and new moon, respectively). As the tides drop, the moon empties (crescent) until the moon is high in the sky during dusk or dawn, at which time the tides fall and the moon runs out of water (first and last quarter). Warner (1937:368) claims that "the Murngin [another name for the Yolngu of Arnhem Land] have a most accurate knowledge of the locational, seasonal, and daily variation of the tides. Anyone who has taken a canoe trip with them along the seacoast quickly learns that this knowledge is immense in detail, well organised, and held by all the men." Warner subsequently describes the important role of the tides, Moon, and Sun in the Yolngu ceremonies and rituals. Tidal data from Milner Bay (Groote Eylandt) and Gove Harbour (Arnhem Land) show that semi-diurnal ranges reach their maximum during the period of full and new moon in coastal areas of the Northern Territory (Northern Territory Transport Group, 2011; see Figure 2).

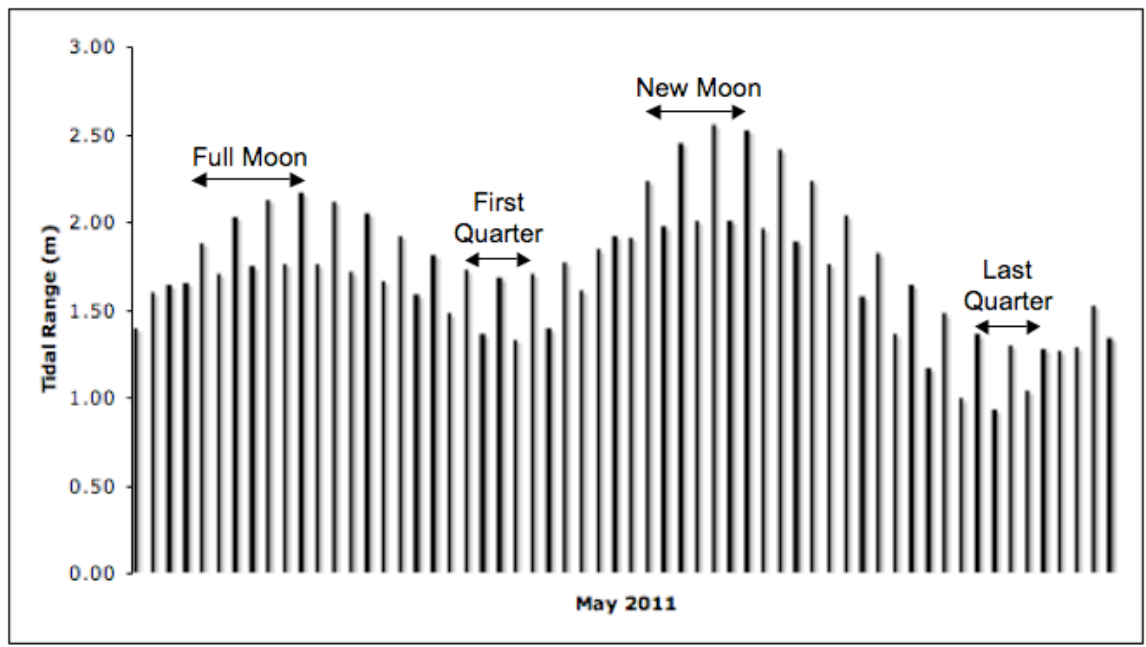

*Figure 2: The tidal range (difference between high and low tide) over the course of May 2011 in Gove Harbour, eastern Arnhem Land, showing the data range that corresponds to particular lunar phases (the day of and two days following). Data taken from Northern Territory Transport Group (2011).*

In addition to describing the lunar phases and their relationship to tides, some Aboriginal groups identified that the earth was finite in expanse. The Yolngu tell how the sun-woman Walu lights a small fire each morning, which we see as the dawn (Wells, 1964). She decorates herself with red ochre, some of which spills onto the clouds, creating the red sunrise. She then lights her torch, made from a stringy-bark tree, and carries it across the sky from east to west, creating daylight. Upon reaching the western horizon, she extinguishes her torch and starts the long journey underground back to the morning camp in the east. When asked about this journey, a Yolngu man told Warner (1937:328) that "the Sun goes clear around the world", who illustrated this "by putting his hand over a box and under it and around again". Smith (1970:93) notes that some Aboriginal astronomers (elders who studied the motions and positions of celestial objects) seem to know the earth was round, as a particular reference to a "day" meant "the earth has turned itself about", although the degree of cultural contamination by Westerners, if any, is uncertain.

These accounts reveal that Aboriginal people were well aware of the motions of the sun and moon and some coastal groups were aware of their correlation with ocean tides. Understanding this relationship is a step to determining the causes of eclipses.

## 4.0    Aboriginal Reactions to Eclipses

### 4.1    Solar Eclipses

Much like other transient celestial phenomena, such as comets and meteors (e.g. Hamacher & Norris, 2011, 2010), many Aboriginal groups held a negative view of solar eclipses. They could be a warning of a terrible calamity, an omen of death and disease, or a sign that someone was working black magic (Wood, 1870:94; Mudrooroo, 1994:59). According to colonist accounts, solar eclipses caused reactions of fear and anxiety to many Aboriginal people, including Aboriginal people near Ooldea, South Australia (Clarke, 1990; Bates, 1944:211), the Euahlayi of New South Wales (Parker, 1905:139-140), the Yircla Meening of Eucla, Western Australia (Curr, 1886:400), the Bindel of Townsville, Queensland (Morrill, 1964:61), the Wirangu of Ceduna, South Australia (Bates, 1944:211), the Ngadjuri of the Flinders Ranges, South Australia (Tindale, 1937:149-151), the Arrernte and Luritja of the Central Desert (Strehlow, 1907:19; Spencer & Gillen, 1899:566), the Kurnai of southeast Victoria (Massola, 1968:162), the people of Roebuck Bay, Western Australia (Peggs, 1903:358, 360) and Erldunda, Northern Territory (Hill, 2002:88). One colonist noted seeing Aboriginal people run under the cover of bushes in a fearful panic upon a solar eclipse (Curr, 1886:400). In 1934, Aboriginal informants of the Mandjindja language in the Western Desert told (Tindale, 2005:361-362) that they called a solar eclipse "Tindu korari", an event they claim to have only seen once. They were struck with great fear at first, but were relieved when the eclipse passed with no harm having come to anyone. Tindale attributed this to an annular eclipse that occurred on 30 July 1916 (Event #1). The most recent annular eclipse visible from this region occurred 246 years prior, while the most recent total solar eclipse occurred 1082 years prior, although four partial eclipses that covered more than 80% of the sun's area occurred between 1900 and 1934 from this region (in 1900, 1905, 1915, and 1922). Although the specific eclipse the Mandjindja witnessed is uncertain, the annular eclipse of 1916 is the best candidate, as it covered 92.4% of the sun's surface.

To some Aboriginal communities of southeast Australia, the sky world was suspended above the heads of the people by trees, ropes, spirits, or magical means. In Euahlayi oral traditions, the sun is a woman named Yhi who falls in love with the moon man, Bahloo. Bahloo has no interest in Yhi and constantly tries to avoid her. As the sun and moon move across the sky over the lunar cycle, Yhi chases Bahloo telling the spirits who hold the sky up that if they let him escape, she will cast down the spirit who sits in the sky holding the ends of the ropes and the sky-world will fall, hurling the world into everlasting darkness (Parker, 1905:139-140).

To combat this omen of evil, some communities employed a brave and well-respected member of the community, such as a medicine man or elder, to use magical means to fight the evil of the eclipse. This typically included throwing sacred objects at the sun whilst chanting a particular song or set of words. This practice was common to Aboriginal communities across Australia, including the Euahlayi, whose medicine men (wirreenuns) chanted a particular set of words (*ibid*) and the Ngadjuri who threw boomerangs in each cardinal direction to avert the evil (Tindale, 1937:149-151). Similarly, medicine men of Arrernte[1] and Pitjantjatjara communities would project sacred stones at the eclipsing sun whilst chanting a particular song – always with success (Spencer & Gillen, 1899:566; Rose, 1957:146-147). The act of casting magical stones at the sun strengthened the medicine man's status in the community since he was always successful in bringing the sun back from the darkness, averting the evil and saving the people. A nearly identical practice was performed in the event of a comet, which yielded the same result (Hamacher & Norris, 2011). Among the Wardaman of the Northern Territory, the head of the sun-clan is a man named *Djinboon*. He can prevent or rescue the earth from an eclipse of the sun by magical means, or allow it to occur and frighten the people if laws are broken or if he does not receive the gifts he desires (Harney & Elkin, 1968:167).

Hill (2002:88) explains that the Aboriginal people near Erldunda, Northern Territory reacted with a combination of fear and joy to a solar eclipse that occurred on 21 September 1922 (Event #2), with some calling out "*jackia jackia*" while others sang, in a fearful tone, the song "*You want to know what is my prize*". However, not all Aboriginal communities viewed solar eclipses with fear, as the Aboriginal people of Beagle Bay, Western Australia were apparently unafraid of solar eclipses (Peggs, 1903:340-341).

### 4.2    Lunar Eclipses

Reactions to lunar eclipses are similar to those of solar eclipses. The Kurnai of Victoria saw a lunar eclipse as a signal that someone they knew on a journey had been killed (Massola, 1968:163). Similarly, Mudrooroo (1994:58) explains that a lunar eclipse was an omen that someone on a journey had a serious accident, although he does not cite a specific Aboriginal group. The Ngarrindjeri near the mouth of the Murray River were fearful of the lunar eclipse of 13 August 1859 (Event #3), believing it to have been created by powerful Aboriginal sorcerers living beyond the European colonial areas (Clarke, 1997:139; Taplin, 1859:2 Sept 1859). Aboriginal people in the Wellington District of Queensland believed a lunar eclipse to be an omen of calamity to a distant relative and reacted with fear and sorrow (Lang 1847:460).

The perception that a lunar eclipse was an omen of death was shared by the Aboriginal people of Beagle Bay, Western Australia. During a lunar eclipse on 23 June 1899 (Event #4), an Aboriginal informant explained to Peggs (1903:340-341) that the eclipse was an omen of death to a man - if the moon is hungry and "wants to eat someone (a man)", it becomes dark – but is uninterested in eating a woman[2]. On the same night, an Aboriginal man from a nearby community told Peggs that among his people, a lunar eclipse represented a man who had become sick.

A Wuradjeri account of a dying "clever man" is associated with what is possibly a partial lunar eclipse. As the man lay dying, 30 km away a corroboree was being held. When some of the people in the corroboree looked up at the moon, they saw the man's *warangun* (spirit) strike the moon, followed by two dark patches that began to cover the moon, which was high in the sky. The people in a corroboree stopped singing and dancing, realising that a lunar eclipse was an omen that someone had died. The next morning, they received the message that the Clever Man had died during the night. He had been lying on his back looking at the moon when he died – at the exact moment the people in the corroboree saw the moon go completely dark (Berndt, 1947/48:83).

In western Queensland, a colonist at Wymullah Station on the Widgeewoggera River recounted a first-hand story about how he exploited a lunar eclipse to reclaim horses stolen by a local Aboriginal group (McNeile, 1903). One day, his horses disappeared and he had reason to believe it was a local group of Aboriginal people. After failing to locate the horses, McNeile approached an Aboriginal man named Jimmy, who requested a ransom of rum, tobacco and clothes in exchange for the location of the horses. Later that day, McNeile read in the local newspaper that a lunar eclipse was predicted to occur that night (20 December, year not given). Using the event to his advantage, McNeile told Jimmy that if he didn't reveal where the horses were, he would make the moon disappear that night. And if they were not returned by the next morning, he would make the sun disappear the next day – permanently. After being ignored by Jimmy, McNeile took a pair of bootjacks, went to the Aboriginal camp, and began dancing and chanting a song in Latin (which he improvised on the spot). As he did this, the people watched and laughed in amusement until the moon began to go dark, which caused confusion and anxiety. As it reached full eclipse, panic struck the people at the camp and they began screaming and running into their huts. The next morning, he found his horses in a nearby small pin. Jimmy informed him that the "Horses found themselves… You no put out big feller sun now, boss? You leave 'm all right?" We attempted to identify a corresponding eclipse using Espenak & O'Byrne (2007b) from various vantage points across Queensland. We failed to identify any lunar eclipses on 20 December between 1800-1903 in any area of Queensland, suggesting the account was simply a fabricated story and not based on an actual event. The similarity of the eclipse story to Mark Twain's novel <u>A Connecticut Yankee in King Arthur's Court</u>, published a few years earlier in 1889, suggests that McNeile's story was simply fiction.

Although many groups viewed lunar eclipses as bad omens, the Aboriginal people near Ooldea, South Australia held no negative views of lunar eclipses, which they called "*pira korari*". They had witnessed one at Wynbring after colonists had built the Transcontinental Railway and paid little attention to it, according to Tindale (1934:21-27). The Transcontinental (or Trans-Australian) Railway was completed in October 1917. In that year, there were three total lunar eclipses visible from this region, suggesting the men witnessed the eclipse on 28 December 1917, which was already eclipsing as it rose above the horizon (Event #5). The frequency of total lunar eclipses visible that year (on 8 January, 5 July, and 28 December) may explain why the event was downplayed by Tindale's Aboriginal informants.

In hunter-gather societies, the sharing of food is essential to the survival of the community and stealing or hoarding food is taboo. The Lardil of Mornington Island viewed the moon as a greedy and selfish man who steals food and gorges, getting fatter (waxing moon). As punishment for this action, he is cut into pieces, getting thinner (waning moon) until he dies (new moon). The sudden and apparent "death" of the moon during a lunar eclipse (McNight, 2005:xxii) served as a mnemonic and warning to younger generations about the moon's selfish nature, reinforcing the taboo of food theft and gluttony.

## 5.0    Causes of Eclipses: An Aboriginal Perspective

### 5.1    Solar Eclipses

From the following accounts, it seems many Aboriginal groups had a firm understanding that during a solar eclipse, an object was covering the sun, although many explanations were presented as to what that object was and why it covered the sun.  However, these explanations were dependent upon the person recording and translating these descriptions, which were nearly always non-Aboriginal people, typically recorded as a passing observation with little detail provided to the reader.

We first present cases where the people understood the moon was the object covering the sun.  In Euahlayi culture, the sun woman, Yhi, is constantly pursuing the moon man Bahloo, who has rejected her advances.  Sometimes Yhi eclipsed Bahloo, trying to kill him in a jealous rage.  However, the spirits that held up the sky intervened and drove Yhi away from Bahloo (Parker, 1905:139-140; Reed, 1965:130).  The Yolngu people of Elcho Island in Arnhem Land provided a similar, but less malevolent, explanation for a solar eclipse: it was an act of copulation between the sun woman and moon man (Warner, 1937:538).  The Wirangu of South Australia believed the solar eclipse on 21 September 1922 (Event #6) was caused by the hand of *maamu-waddi*, a spirit man that covered the earth during the eclipse for the privacy of the sun woman and moon man while they were *guri-arra* – "husband and wife together" (Bates, 1944:211).  Near Eucla, South Australia, the Yircla Meening believed solar eclipses were caused by "the Meenings of the moon, who were sick, and in a bad frame of mind towards those of Yircla[3]" (Curr, 1886:400).  This account implies a link between the moon and sun during an eclipse, although the cause is not specifically stated.  In these cases, it is clear the Aboriginal people understood that the moon covered the sun during the eclipse (except for the latter account which is ambiguous).

Such an understanding suggests that the Aboriginal group were aware of the moon's position in the sky through its various phases.  Despite the fact that the moon is essentially invisible for three days during the period of "new moon", an observer who had been following the position of the Moon throughout the month would be able to predict its position during the New Moon phase.

Among other communities, it is clear that the people understand *something* was covering the sun during a solar eclipse, but attributed that "something" to various objects or actions, including a large black bird called *tia* to the Arrernte (Strehlow, 1907:19) or spun possum fur to the Luritja (*ibid*). To some Aboriginal groups in southwestern region of Western Australia, a solar eclipse is caused by *mulgarguttuk* (sorcerers) placing their *booka* (cloaks) over the sun, while to some other groups they move hills and mountains to cover the sun (Bates, 1985:232). A similar view is held by Aboriginal people of the Central Desert who call a solar eclipse *bira waldurning* and claim it is made by a man (*waddingga*) covering the sun with his hand or body (Bates, 1904-1912: Notebook 6a, p. 74). During an eclipse of the sun on 5 April 1856 (Event #7), a Bindel man told Morrill (1864:61) that his son covered the sun and caused the eclipse in order to frighten another person in the community. An earlier Arrernte account attributes a solar eclipse (*Ilpuma*) to periodic visits of the evil spirit *Arungquilta* who takes up residence in the sun, causing it to turn dark (Spencer & Gillen, 1899:566). The Pitjantjatjara of the Central Desert believed that bad spirits made the sun "dirty" during a solar eclipse (Rose, 1957:146-147) while the Wardaman believed a solar eclipse was caused by an evil spirit swallowing the sun (Harney & Elkin, 1968:167). The Wheelman people of Bremer Bay, Western Australia told Hassell & Davidson (1934:234-236) a story about how one day the sun and moon fell to earth, splitting it in half. The lazy people were separated from the rest of the community to the other side of the sun. Sometimes they got bored and wanted to see what was happening in this world. As they tipped the sun on its side to have a peek, several of them would gather, blocking the sun's light, causing a solar eclipse. They only do this for a short time – just long enough for each of them to have a look, which explains why the eclipse does not last long. Hassell's informant told her that "Yhi (the sun) hide him face and Nunghar look down," when storms come or the sky becomes dark in the daytime (solar eclipse). These accounts reveal an understanding that an object is covering the sun during an eclipse, whether it is by natural or magical means, although the obscuration is not attributed to the moon.

Not all causes of solar eclipses were attributed to an object covering the sun. According to a community in Turner Point, Arnhem Land, a solar eclipse was caused when a sacred tree at a totemic site was damaged by fire or carelessness (Chaseling, 1957:163). As such, sitting under the tree or even seeing it is reserved solely for initiated elders. One final account provides no insight to the cause of the eclipse, but provides an interesting account of how tangible and nearby some Aboriginal people thought the sun to be. When astronomers in Goondiwindi, Queensland were observing and recording the total solar eclipse of 21 September 1922 in order to test Einstein's General Theory of Relativity, some Aboriginal people present thought the astronomers were trying to catch the sun in a net (Menzel, 1949:275, Event #8). Unfortunately, the Menzel gives no further information as to why the Aboriginal people thought this or to their reactions during or after the eclipse.

### 5.2    Lunar Eclipses

We only find a few accounts describing the causes of lunar eclipses. Reed (1965:130) describes the story (repeated by Johnson, 1998) that the moon-man is constantly pursued by the sun-woman and manages to avoid her advances most of the time. However, he is occasionally overtaken by the sun-woman, signified by a lunar eclipse. Although Reed does not cite the Aboriginal group or location from which the story was taken, it is very similar to the Euahlayi account of a solar eclipse. The Arrernte believed a lunar eclipse was the result of the moon man hiding his face behind the possum fur that he is constantly spinning, which is identical to the Luritja view of a solar eclipse (Strehlow, 1907). As with a solar eclipse, Aboriginal groups in southwest region of Western Australia believe a lunar eclipse is caused by *mulgarguttuk* placing their cloaks or a hill/mountain over the moon (Bates, 1985:232). The Kayardild of Bentinck Island in the southern Gulf of Carpentaria believed the moon was a man who used a net (halo of the moon) to collect the souls of recently dead during a lunar eclipse (*jawaaja*). As the net filled, the moon-man would disappear, as if he himself had died, which prompted the people to hide under fig trees, fearful that the moon would kill them. If the people did not seek shelter, they would be struck with *jiljawatha*, a sickness that induced crusted sores (Evans, 1995:590-596). Róheim (1971:53) suggests a Eucla Dreaming that describes a man ascending to the Milky Way who can only bee seen when he "walks across the moon" may describe a lunar eclipse, showing an understanding that an "object" (the earth's shadow) covers the moon during lunar eclipses.

The generally reddish colour of the moon observed during a total lunar eclipse, as discussed in Section 2, is noted by some Aboriginal groups, including the Aboriginal people of the Clarence River, New South Wales who thought a lunar eclipse revealed the moon-man's blood (Mathews, 1994:60) and the Kurnai of Victoria who believed a red moon signified that someone had been killed (Massola 1968:162). The Lardil of Mornington Island believe the moon man's blood is visible during a total lunar eclipse, prompting elder people to shout out "don't kill him!" (McNight, 1999:105). Strehlow (1907) notes that the Luritja believed the moon sometimes goes into the graves of the recently dead and eats the entrails of the bodies. He then emerges into the sky, blood red in colour, so everyone can see what he has done. However, Strehlow claims this account has nothing to do with lunar eclipses but is instead referring to the new moon. The moon can take on a reddish hue when it is low on the horizon, because the shorter wavelengths of light are reduced as they pass through the atmosphere at a low angle, allowing the longer wavelengths to dominate the colour.

In Ungarinyin culture of Western Australia, an unfriendly medicine man causes the face of the moon to be covered with blood, which greatly frightens the people (the text is unclear how this is done, but is presumably by some magical means). A friendly medicine man then ascends into the sky during a dream. Upon his return, he informs the people that he made the moon "better" (Elkin, 1977:126)[4].

## 6.0     Dating Oral Traditions from Historic Eclipses

The age of a story that includes a description of a natural event may be estimated by identifying the date of that event.  Tindale (1937:149-151) believed that a Ngadjuri story from Parachilna, South Australia described a solar eclipse, which he dates to 1793.  In the story, an elderly female being came from the northwest accompanied by two dingoes who behaved as men - one with reddish fur and the other with black fur.  Two brothers, *Wulkinara* and *Kudnu* of the lizard totem, succeeded in killing the dingoes and burning the old woman.  As a result, the sun disappeared, causing fear among the people.  The community tried diligently to bring the sun back from the darkness, eventually collapsing, exhausted and in tears, and fell asleep.  Kudnu awakened during the darkness and cast magic boomerangs into the sky in each of the cardinal directions. The first three - to the north, south, and west - failed, but the fourth, cast towards the east, was successful and the sun appeared again.   Tindale attributes this event to a total solar eclipse that passed over Parachilna on 13 March 1793 (see Figure 3).  Using Espenak and O'Byrne (2007a) and the Starry Night astronomical software package, we calculated that the solar eclipse that passed over Parachilna actually occurred a day earlier on 12 March 1793, and was a partial eclipse that covered ~93% of the sun's area (Event #9a) with the path of totality passing over 200 km north of Parachilna, although Tindale may have considered what was visible from Parachilna as "total".  Tindale claims that any earlier (total) solar eclipses would have occurred prior to 1600.   However, there was a total eclipse visible from Parachilna just eleven years prior on 7 October 1782 (Event #9b). Although there were no other total eclipses visible from Parachilna between 1701 and 1782, there were six other partial eclipses that covered 75% or more of the sun during that time.  The most recent total solar eclipses prior to 1793, aside from the 1782 event and a few annular eclipses, were in 1608 and 1610.   A better candidate for Tindale's explanation would be the total solar eclipse in October 1782 that passed over Parachilna. As we are unclear of Tindale's definition of a "total eclipse", there will be some ambiguity in this interpretation.  In areas where the 1793 total eclipse would have been visible, such as at Lake Eyre, the planet Mercury would have become clearly visible just 1.5 degrees above the sun during totality, but there is no mention of this in the story.

However, the story described the woman as coming from the northwest and the 1793 eclipse was visible in the west-northwest sky, while the 1782 eclipse was visible in the east-northeast sky.

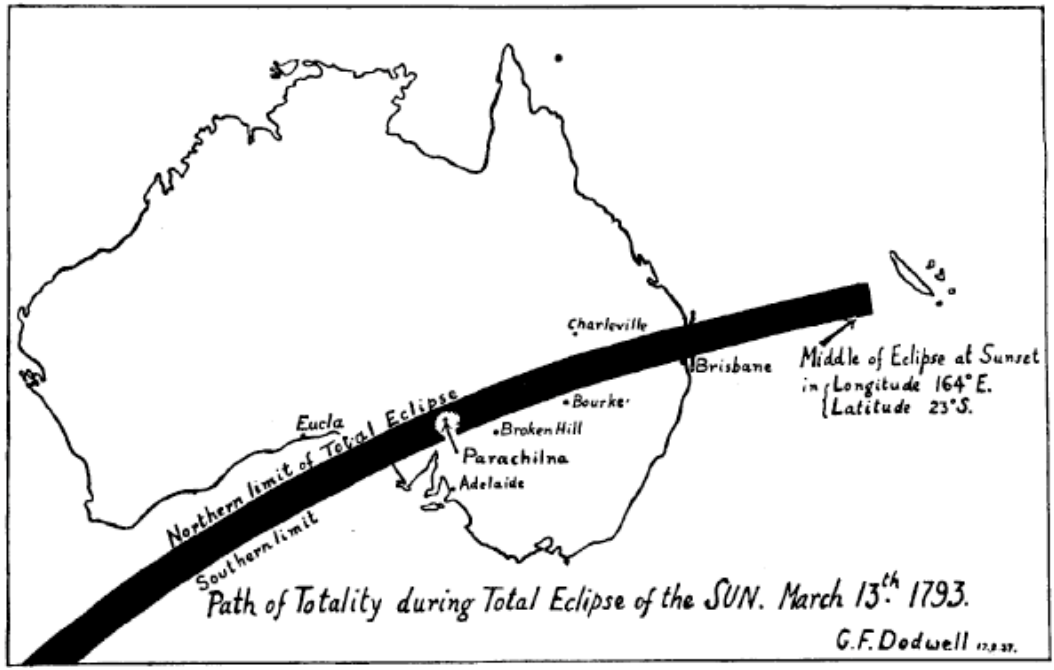

*Figure 3: The path of a total solar eclipse that occurred on 12 March 1793 as calculated by Government Astronomer Mr. G.F. Dodwell that Norman Tindale believed was the source of an Aboriginal story about the sun becoming dark (Dodwell's calculation was off by a day and the total eclipse did not cover Parachilna). Image taken from Tindale (1937:152).*

A problem arises with Tindale's interpretation of this story as representing a solar eclipse: the story describes the people falling asleep while the sun goes dark and waking sometime later with the sun still dark. Under the best conditions, the sun will remain in totality (completely covered) for no more than 7.5 minutes. The total duration of the 1782 eclipse was ~2.5 hours, with totality lasting only ~2.5 minutes. The people would have only been in total darkness for only a couple of minutes – not long enough to exhaust one's self into sleep then awake sometime later still in darkness. Another explanation would have been heavy cloud-cover, although it seems unlikely people would react in such a fearful panic to mere clouds.

### 7.0     Aboriginal Prediction of Eclipses

To understand the cause of eclipses is to understand the relationship between the motions of the sun and moon over time.  If these motions are understood with sufficient accuracy, an eclipse can be predicted in advance.  However, the required accuracy is very high, requiring carefully constructed instruments to make the required measurements. We found only one account that mentioned such a prediction: Peggs (1903:358, 360) presents letters written by her to C.J. Tabor whist Peggs was living in Roebuck Bay, Western Australia between 1898-1901.  Peggs (1903:358) wrote "We are to witness an eclipse of the sun next month. Strange! all the natives know about it; how, we can't imagine!" (letter dated 'December 1899').  Peggs asked a local Aboriginal woman named Mary about the eclipse, who responded "Him go out all right".  It is unclear from her account how she concluded that Mary had predicted the event – whether it was Mary's comment or by some means not described in the letter.  The comment by Mary, however, may have been misleading, as she may have merely been acknowledging what happens during an eclipse. Peggs later wrote, "The eclipse came off, to the fear of many of the natives.  It was a glorious afternoon; I used smoked glasses, but could see with the naked eye quite distinctly.  There seemed such a rosy hue surrounding the sun, at times changing to yellow.  After a good deal of persuasion Jack convinced Mary to look through glasses, but she was half afraid."  Given that the letter was dated December 1899, we searched for any solar eclipses during this period.  Between 1891 and 1900, only one solar eclipse was visible from this region, a partial eclipse that covered 73% of the suns surface, which occurred on 22 November 1900 (Event #10).

Reasons for doubting the veracity of this story include (a) the inconsistency in the dates, (b) the lack of evidence that Aboriginal people made sub-arcminute precision measurements required for eclipse prediction, despite evidence elsewhere for Aboriginal astronomical alignments accurate to a few degrees (e.g. Wurdi Youang, Norris et al, 2011; stone rows, Hamacher & Norris, 2011), and (c) a reaction of fear to something they would have predicted seems counterintuitive.

**8.0     Representations of Eclipses in Aboriginal Rock Art**

Astronomical symbolism is found in Aboriginal rock art across Australia (see Norris & Hamacher, 2011).   Ku-ring-gai Chase National Park, north of Sydney, is home to a number of Aboriginal rock engravings, some of which depict crescent motifs (see Figure 4).    Traditionally, archaeologists (e.g. McCarthy, 1983) refer to these motifs as boomerangs.    However, we are currently conducting a detailed study to determine, statistically, if these shapes more likely represent crescent moons, boomerangs, or an eclipsing sun.  An engraving at Basin Track in Ku-ring-gai Chase National Park depicts a man and woman, their arms and legs overlapping, with a crescent shape above their heads.   While other engravings depicting a man and woman partially superimposed are found in the region, with only their arms and legs intersecting, see McCarthy, 1983) the crescent above their heads is found only at Basin Track.   In other cases, the male figure is holding a crescent in one hand and a fish or shield in the other and some engravings show a single figure with a crescent above the head.  The meaning of these motifs is unclear, as few ethnographic records regarding these engravings and ceremonial sites exist.   In the case of the Basic Track engraving, Norris & Norris (2009) speculated that the motif might represent the moon-man obscuring the sun-woman during a solar eclipse[5].   Near the man-woman is an engraving of a hermaphrodite figure, which could represent the sun and moon in full eclipse (John Clegg, 2009 - personal communication).  If we speculate that this motif represents a solar eclipse as seen from that location in the direction of the engraving (i.e. a straight line from the feet of the figures through their heads and crescent, towards the horizon), the eclipse must occur near dawn, as the petroglyph faces 55±5 degrees east of north.  We examined solar eclipse events visible from the region in the 19[th] century using the Starry Night software package.  One eclipse candidate occurred at dawn on 8 August 1831 ($t_{start}$ = 06:45, $t_{max}$ = 07:03, $t_{end}$ = 08:13), which covered ~85% of the sun's surface (Event #11).   At mid-eclipse, the sun closely resembles the crescent engraving, with the cusps of the crescent pointing downward (see Figure 5).   The engraving aligns to the general direction the eclipse would have been visible from this location (between due east and 45 degrees northeast).   Unfortunately, we have no supporting ethnographic evidence and dating a rock engraving is problematic, as engravings were typically re-grooved during ceremonies (Stanbury & Clegg, 1990). Therefore, this interpretation remains speculative.

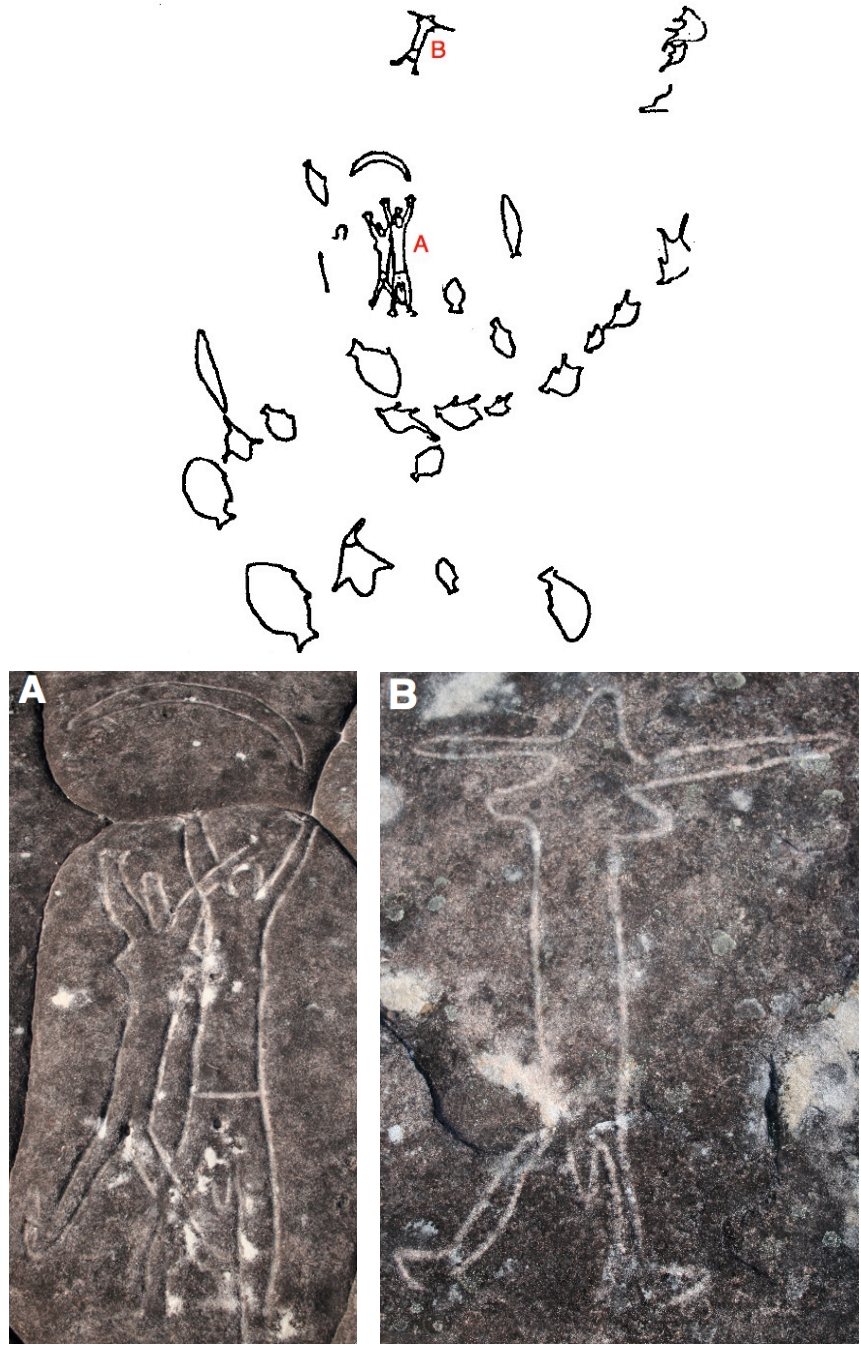

*Figure 4: (Top) Aboriginal rock engravings at the Basin Head Track, Kuringai Chase National Park, taken from Stanbury & Clegg (1990), with images of the man and woman engraving (A) and the hermaphrodite figure (B). Photos by D.W. Hamacher.*

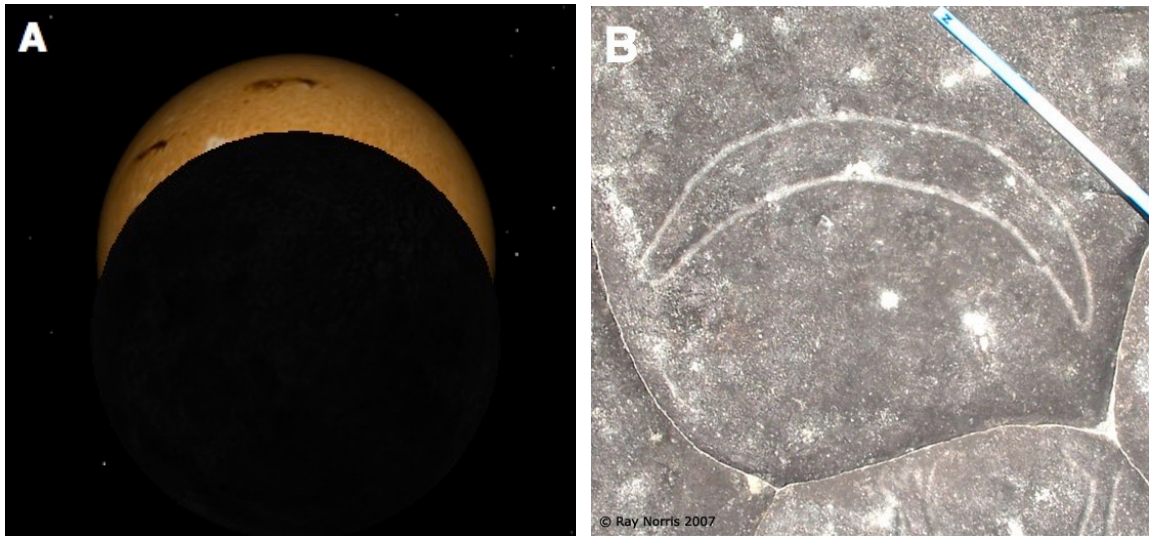

*Figure 5: (A) A partial solar eclipse that occurred on 8 August 1831 as viewed from Kuring-Gai Chase National Park just after 07:00 during mid-eclipse (Event #10), although the exposed sunlight form even this eclipse would still cause retinal damage. Image taken from the Starry Night astronomical software package. (B) The Basin Head Track engraving of a crescent shape above the heads of a man and woman who are partially superimposed. The blue and white stick shows the orientation of magnetic north, which is 12.5 degrees east of true north from this location (Image by R.P. Norris).*

## 9.0    Discussion and Conclusion

Given the low probability of witnessing a total solar eclipse in Australia, we expected to find very few accounts of total solar eclipses.  And since a partial eclipses can pass without notice because of the sun's intense brightness, and because of the damage to the eye that can result from directly looking into the sun, we did not expect to find many accounts of partial eclipses, either.  Of the four accounts that we can attribute to a specific solar eclipse, three of them are partial eclipses, with some obscuring as little as 75.7% of the sun's surface.  We also find a number of Aboriginal words and descriptions of solar eclipses, despite our initial predictions.  This shows that Aboriginal people did observe some total and partial eclipses and the memory of these events remained strong in many areas.  We cannot attribute any partial eclipses that covered less than 75% of the sun's surface to oral traditions and would use this as an estimated lower limit to what people could reasonably notice, although observing the sun even when 75% is eclipsed would still cause retinal damage.  However, we acknowledge that other factors can reveal partial eclipses, such as diffraction by tree leaves, sufficient cloud-cover, or low-horizon partial eclipses, where the intensity of the sun's light is reduced (Figure 6).

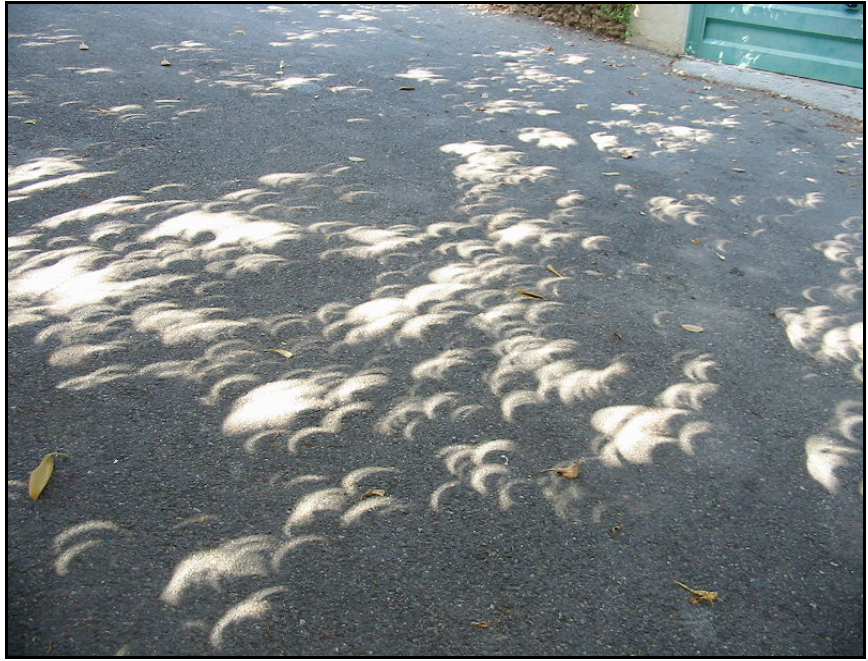

*Figure 6: Tree leaves acting as pinholes on 3 October 2005 in St. Juliens, Malta, allowing one to observe a partial eclipse without looking at the sun.  Image reproduced under Wikipedia Commons license.*

The available data reveal that some Aboriginal groups understood the mechanics of the sun-earth-moon system and the relationship of lunar phases to events on the earth.  The Yolngu people of Arnhem Land provide the most complete ethnographic evidence, in that their oral accounts demonstrate that they understood that the sun and moon move in an east to west motion, the moon goes through repeated phases that affect the ocean tides, the earth is finite in space, and the moon covers the sun during a solar eclipse.

Particularly important are the accounts that Aboriginal people understood that lunar eclipses were associated with the Sun (Reed, 1965; Johnson, 1998).  It is not surprising that someone familiar with the relative motion of the Sun and Moon might notice that a solar eclipse occurs when the moon is close to the sun, and deduce that a solar eclipse was caused by the moon.  But it would be an impressive intellectual feat for an individual to recognise that a lunar eclipse was connected with the position of the Sun.   It is therefore important to get further independent evidence of knowledge of this association from historical accounts, in order to corroborate the account by Reed.

While conducting ethnographic fieldwork in 2006, one of us (Norris) was with a Yolngu ceremonial leader during a lunar eclipse.  The leader (name withheld) told Norris that his clan had no oral tradition about the eclipse.  However, it is possible that the leader did not want to share this information, as it may have been considered sacred and secret.

Overall, the cosmos is predictable, with most changes occurring gradually and slowly, such as the change in stellar positions over the night or throughout the year, the phases of the moon, or the positions of the planets.  The night sky served many important functions and roles within Aboriginal communities, including time-keeping, food economics, navigation, social structure, marriage classes, and as a mnemonic device.  Surprising transient phenomena, such as eclipses, are relatively rare.  This is probably the reason that eclipses are met with reactions of fear and anxiety and why they are generally associated with negative attributes, such as death and disease – a reaction common to other surprising transient phenomena, such as meteorite impacts, fireballs, and comets (see Hamacher & Norris, 2009, 2010, 2011, respectively).  These perceptions are shared by many other cultures of the world.

Some interpretations presented in this paper are solid examples of "Aboriginal Astronomy" in that they clearly display an understanding of the motions of the sun and moon and their relationship with eclipses, including those of the Yolngu, who had a clear understanding that the Moon covered the sun during a solar eclipse.  Other groups, such as the Wirangu, and Euahlayi, understood that *something* was obscuring the Sun during a solar eclipse, although it is not clear whether they defined that object as the moon.

**Notes**

[1] Among the Arrernte (Anglicised as Aranda or Arunta), eclipses are cause by periodic visits of an evil spirit-magic called Arungquilta that takes up residence in the sun. Arungquilta is also found in meteors and comets (Hamacher & Norris 2010, 2011).

[2] He also noted that if a child were born during a lunar eclipse, the child would be a boy.

[3] Yircla was the name of the community (Eucla) and also that of the Morning Star (Venus).

[4] Although Elkin does not identify whether the heavenly object is the sun or moon, we interpret the account to refer to the moon since the moon turns red during a lunar eclipse.

[5] While earlier illustrations of the engraving show the woman covering the man, the engraving itself is less clear. The engraving lines that comprise the arms and legs of the man and woman cross each other with no special reference to superposition.

**Acknowledgements**

The authors would like to acknowledge the Wallumedegal People (the Traditional Custodians of the land on which Macquarie University is situated), Yolngu Elders, NSW Parks & Wildlife, David Frew, John Clegg, Kristina Everett, and the Mitchell Library in Sydney. This research made use of the Starry Night® astronomical software package, the 'Eclipse Explorer', developed by Fred Espenak and Chris O'Byrne (NASA), and Google Earth. Hamacher was funded by the Macquarie University Research Excellence Scholarship.